\documentclass[a4paper, 10pt, twocolumn, superscriptaddress]{revtex4}

\usepackage{graphicx}
\usepackage[utf8]{inputenc}

\begin{document}

\title{Identifying opinion-based groups from survey data: a bipartite network approach
}

\author{P. MacCarron}
	\affiliation{Centre for Social Issues Research, University of Limerick, Limerick, Ireland}
	\affiliation{MACSI (Mathematics Applications Consortium for Science and Industry), University of Limerick, Limerick, Ireland}
	\author{P. J. Maher}
	\affiliation{Centre for Social Issues Research, University of Limerick, Limerick, Ireland}

	\author{M. Quayle}
	\affiliation{Centre for Social Issues Research, University of Limerick, Limerick, Ireland}
	\affiliation{Department of Psychology, School of Applied Human Sciences, University of KwaZulu-Natal, Private Bax X01, Scottsville, South Africa}

\begin{abstract}

A survey can be represented by a bipartite network as it has two types of nodes, participants and items in which participants can only interact with items. We introduce an agreement threshold to take a minimal projection of the participants linked by shared responses in order to identify opinion-based groups. We show that in American National Election Studies- data, this can identify polarisation along political attitudes. 

We also  take a projection of attitudes that are linked by how participants respond to them. This can be used to show which attitudes are commonly held together in different countries or communities. 

\end{abstract}

\maketitle

\section{Introduction}

The concept of opinion-based groups is a relatively new and important development for our understanding of intergroup relations and polarisation from a social-identity perspective~\cite{bliuc2007opinion, mcgarty2009collective}. Opinion-based identification can exacerbate social problems or solve them. For example, addressing urgent collective issues such as the climate emergency or a global pandemic requires widespread solidarity. Solidarity requires the coordination of core opinions (e.g. agreement that climate change is occurring). 

Opinion-based group identification is particularly likely to inspire collective behaviour, because coordination of opinions is an important step towards collective agency~\cite{mcgarty2009collective,Hogg2007a}. However, even in the face of overwhelming evidence, these core opinions often become the focus of partisan alignment where different ``sides'' of the issue become incorporated into partisan group identities~\cite{Bliuc2015,maher2020mapping,van2020using}. Once opinions are absorbed into partisan identities it becomes difficult to achieve the broad consensus required for decisive democratic responses.

Opinion-based identification is an important feature of online interaction, and groups formed around shared opinions can emerge rapidly though online social media interactions~\cite{garcia2019eatlancet}. While geographical proximity or shared experience have historically been drivers of interaction and identification, users of social media transcend these. Instead, interactions often centre around shared (or disputed) interests and attitudes, and these become the basis for perceived similarity and difference~\cite{macy2019opinion}. 
 
Even offline groups form based on shared opinions through general process such as homophily---that we tend to associate with those similar to ourselves~\cite{mcpherson2001birds}---and the need for group differentiation~\cite{jetten1997strength}.  However,  these processes are amplified in online interaction. First, the affordances of online media platforms are oriented to opinion-sharing. Online interaction frequently centres around ``liking,'' ``upvoting/downvoting,'' ``retweeting,'' and other forms of interaction related to attitude sharing and evaluation. Second, the personalisation of information flow on social media  has lead to the emergence of filter bubbles or echo chambers that reinforce opinion-based differentiation~\cite{bakshy2015exposure}. 

Social networks record traces of interaction and information flow. However, people can interact without ever influencing each-other~\cite{belanger2020countermessaging}. Conversely, groups of people in close agreement can influence society without ever interacting, for example when voting in elections. We refer to the network of agreements and disagreements between people on a field of attitudes as a \textit{homophily network}. 

In traditional opinion-dynamics models (for example~\cite{axelrod1997dissemination}), the homophily network can be even more influential on outcomes than the social network~\cite{valori2012reconciling}. In online interactions, it is often the case that people have little influence over each-other if they perceive themselves to be on different ``sides.'' Instead, interaction in these circumstances may even drive opinions further apart~\cite{belanger2020countermessaging}. So while the homophily network does not tell us much about actual interaction, it can reveal a great deal about the structure of social groups and likely paths of social influence~\cite{breiger2014comparative,buabeanu2018ultrametricity,erickson1988relational}.

Several methods exist for visualizing the social structure and belief systems, conceived as an inter-related network of opinions and values~\cite{boutyline2017belief,brandt2019central}. These approaches are invaluable for mapping the opinion-space of a society and identifying how opinions are tied together in a network of inter-related beliefs. However, our interest is particularly in the structure of social groups and group-identities afforded by the opinion space. We want to analyse the way people are bound together or divided by opinions as well as how opinions are tied together.

In the social sciences the most common way of measuring attitudes is through the use of surveys. Survey data is therefore generally more available and easier to obtain than accessing data from closed online platforms. Most usefully, survey data is coded (by participants) at the point of collection, typically by rating a series of statements (or questions) on ordinal scales \cite{DeVellis2003}.


In this paper, we build on earlier proposals that survey data can be treated as a bipartite network~\cite{breiger1974duality,breiger2014comparative}. This enables researchers to capture what Breiger calls the ``duality'' of the opinion space which contains, simultaneously, clusters of individuals who share similar opinions and clusters of opinions shared by similar individuals. 
This can be done by taking projections from the bipartite graph of the survey.  
The groups of participants identified by shared responses can be used to show polarisation in some situations and the attitudes that make up their group identity can be easily viewed. 
There are many methods for taking projections from bipartite networks, for example~\cite{Horvat2013}, here we introduce an agreement threshold to form the groups and reduce the density. 

There are other methods for connecting shared attitudes or belief systems eg. \cite{brandt2019central,dalege2016toward}. These methods tend to use models from statistical physics resulting in partial correlations between shared attitudes, these are then cut-off below a certain correlation and the graphs are displayed. It can, however, be challenging to interpret what an edge between two items represents or how to decide to threshold the model. There are often many parameters involved in these decisions. 
Using the method presented in this paper, the initial visualisations are parameter-free and the projections have a single parameter -- the agreement threshold. This can initially be set to establish a giant component connecting the majority of participants. This is essentially the backbone of the participant network, lowering this threshold then allows further agreement between participants.

\section{Methods}

As a survey consists of a group of participants responding to a number of items, it is, by definition, a bipartite graph (i.e. it has two types of nodes where nodes of one type may only interact with nodes of another type). The responses to an item are generally given on a scale (e.g. a Likert-type scale). The value associated with this can be assigned as a property of each edge (similar to a weight). The idea of representing this kind of data as a bipartite graph is not new, for example this has previously been done linking senators in the US Supreme court to the motions they voted on and linking Netflix users to media they liked or disliked~\cite{mrvar2009partitioning, Horvat2013, wyse2014inferring}. 
Here and in the following sections we take secondary survey data, and display some of their bipartite graphs and corresponding projections. 

\begin{figure*}
	\centering
	\includegraphics[width=0.95\textwidth]{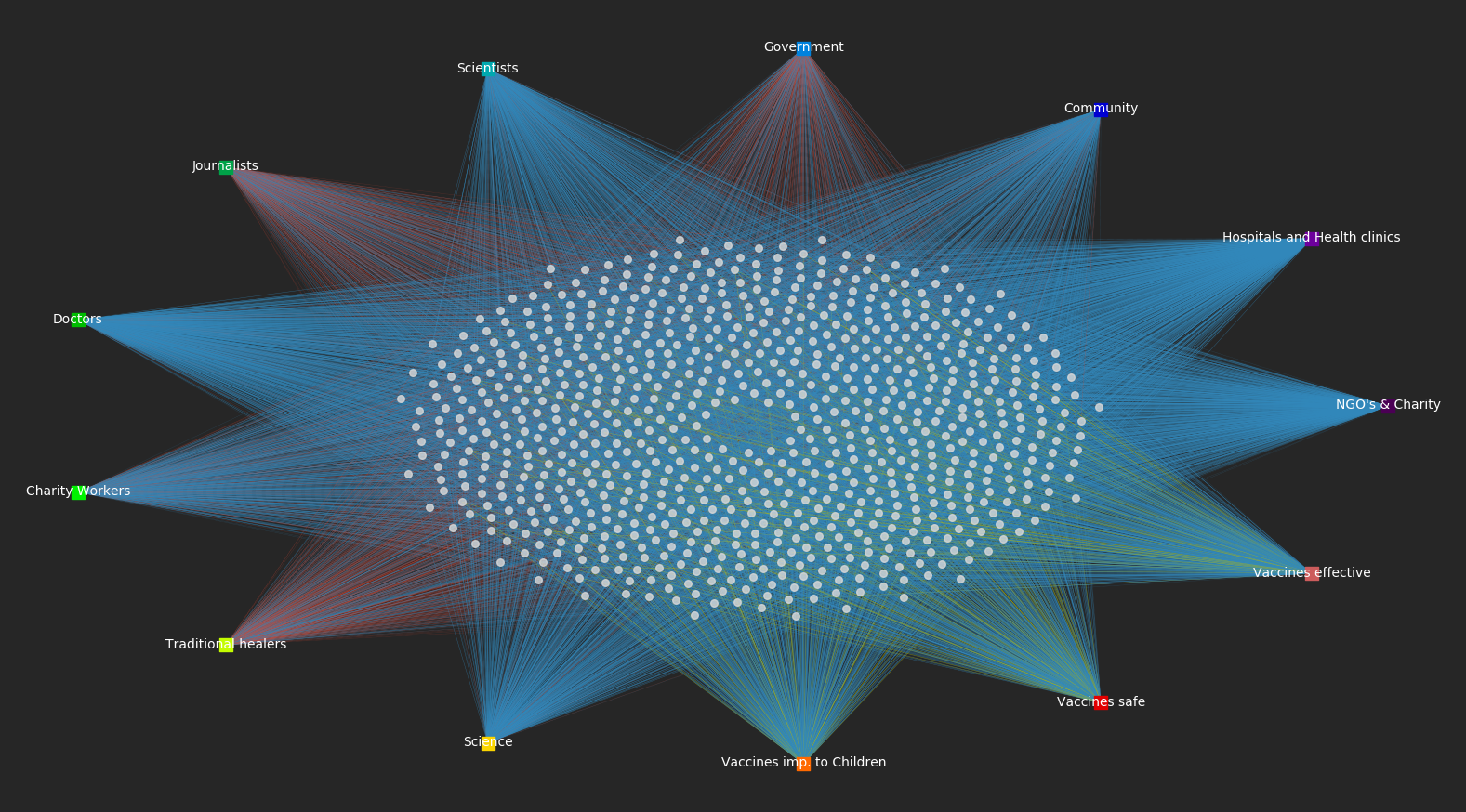} 
	\includegraphics[width=0.95\textwidth]{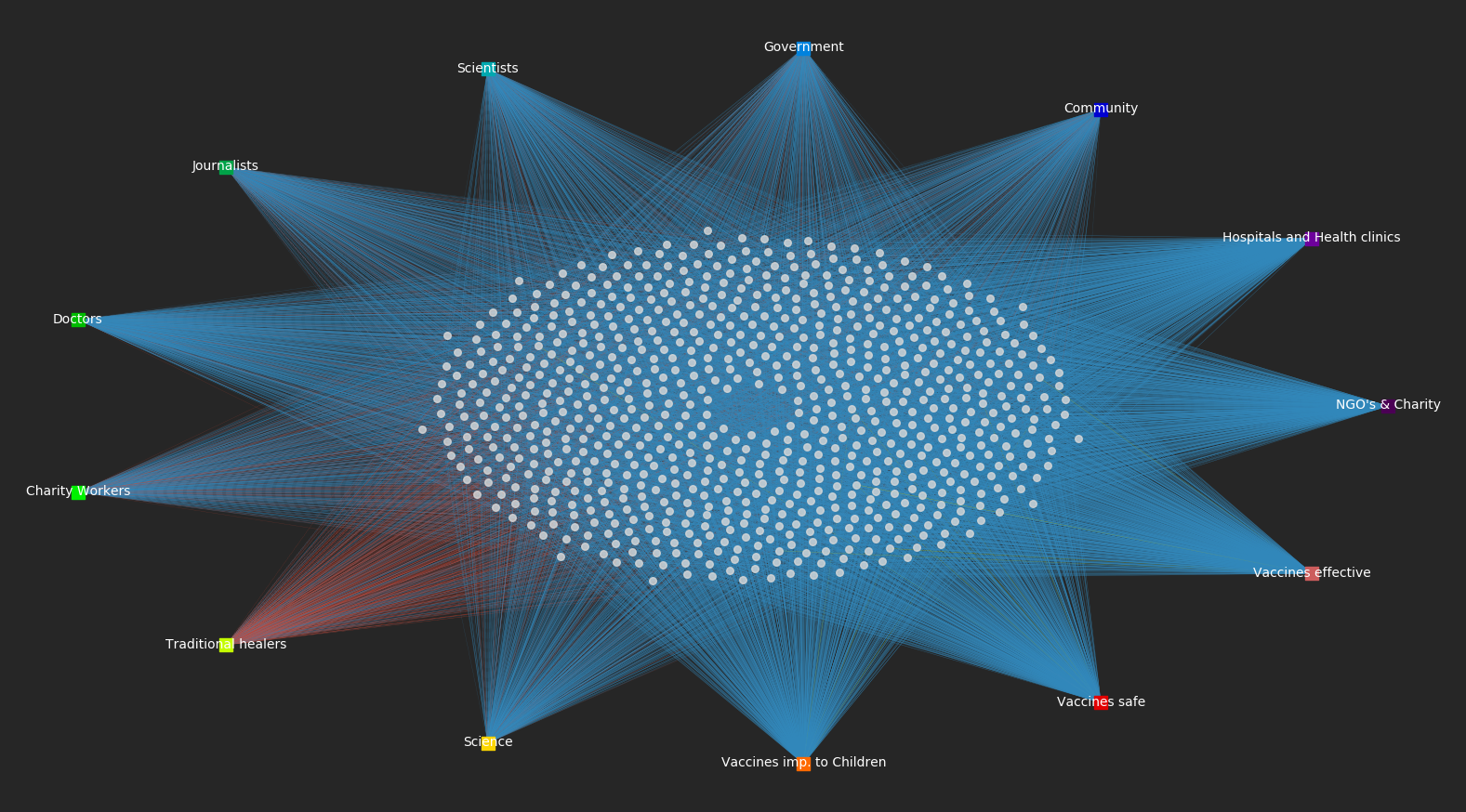} 
	\caption{The bipartite graphs representing the responses to the Wellcome Trust Global Health Monitor 2018  survey for France (top) and Bangladesh (bottom). A red edge represents a participant responding negatively to the item, a yellow is neutral and blue is positive.}
	\label{fig:wellcome_bipartite}
\end{figure*}

The Wellcome Global Monitor 2018 surveyed almost 150,000 people on attitudes towards science and vaccines in 149 countries \cite{monitor2018does}. 
In fig.~\ref{fig:wellcome_bipartite}, we show the bipartite graphs for almost one thousand respondants in France and Bangladesh, two countries identified in the report\cite{monitor2018does} as having differing attitudes towards vaccines. The survey consists of 13 items, 10 having a four-point scales and the three vaccine questions have five-point scales. In fig.~\ref{fig:wellcome_bipartite} we display this as a bipartite graph with France on top and Bangladesh on the bottom. Here, a solid blue [red] edge represents full trust [distrust] in a response, a dashed blue [red] edge represents some trust [distrust] and a yellow edge represents a neutral response (only in the five-point responses).

While these visualisations alone do not tell us too much, we can easily see that many participants in Bangladesh respond positively to all items except Traditional Healers, in France there is significant distrust of the Government as well as some distrust towards 
Journalists. 
There are also more neutral responses to the vaccine questions compared to Bangladesh. 
It is worth pointing out, that at this stage we have lost no information and have done no filtering. 

\subsection{Projections}

We next take projections of the bipartite graph to connect attitudes linked by being co-held by participants, and linking participants with a high agreement on items.

The projections of each of these can be done in different ways. For example, the easiest method is to give a weight to the edge between each participant based on the number of items they give the same response to. That way, an edge with weight 1 between two participants means they answered the same on just one item, whereas, for example, an edge of 12 means they answered the same in 12 of the 13 items. A projected network can be built up by adding edges of highest weight until a giant component (a large subset of nodes where a path can be made between any pair within the component) is formed (this is like the ``backbone'' of the network, a similar method for constructing a network is found in \cite{gleiser2007become}). Once a giant component is formed, the process can be stopped. Therefore we introduce the one and only parameter, the \emph{agreement threshold}, to reach a giant component.

\begin{figure*}
	\centering
	\includegraphics[width=0.95\textwidth]{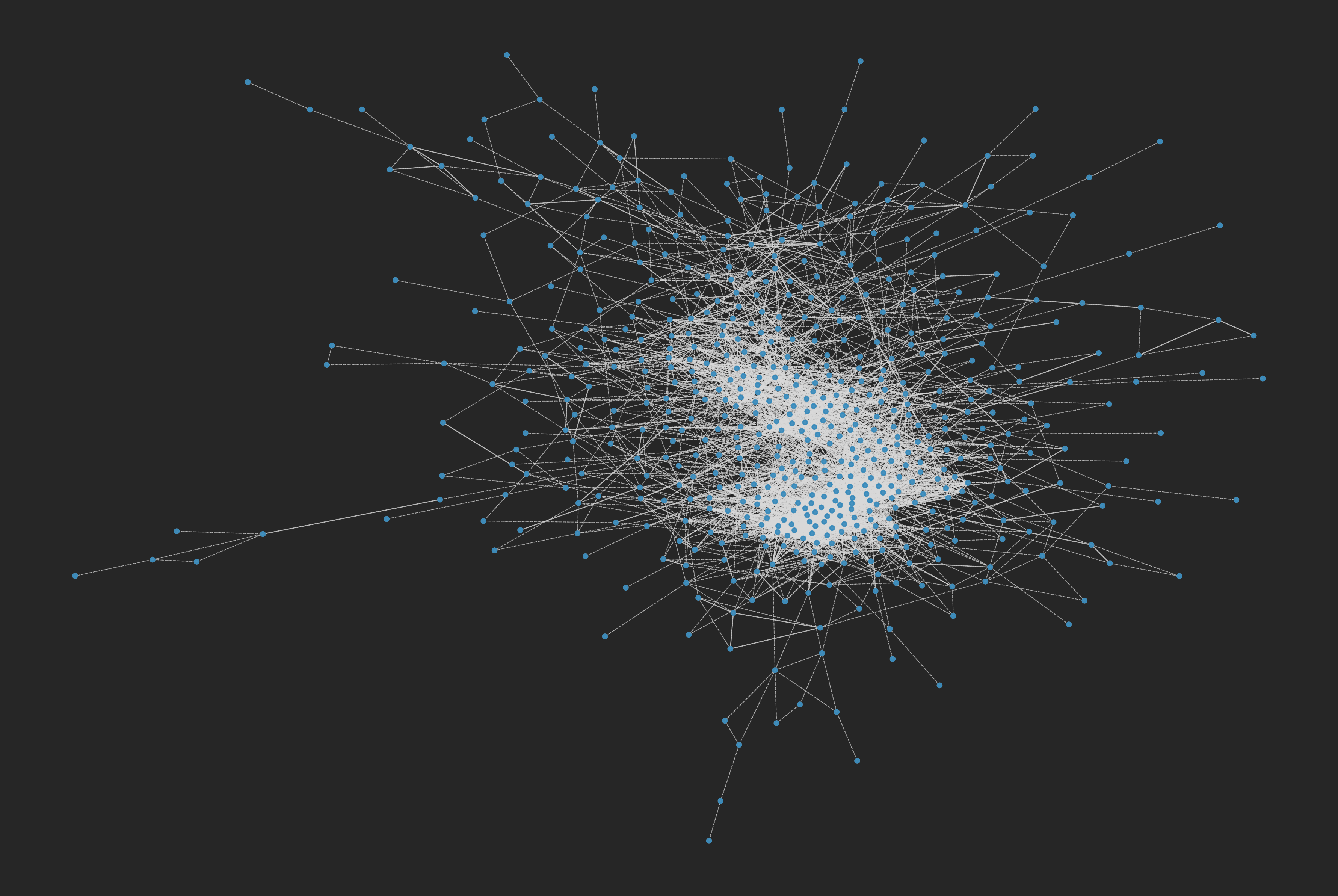} 
	\includegraphics[width=0.95\textwidth]{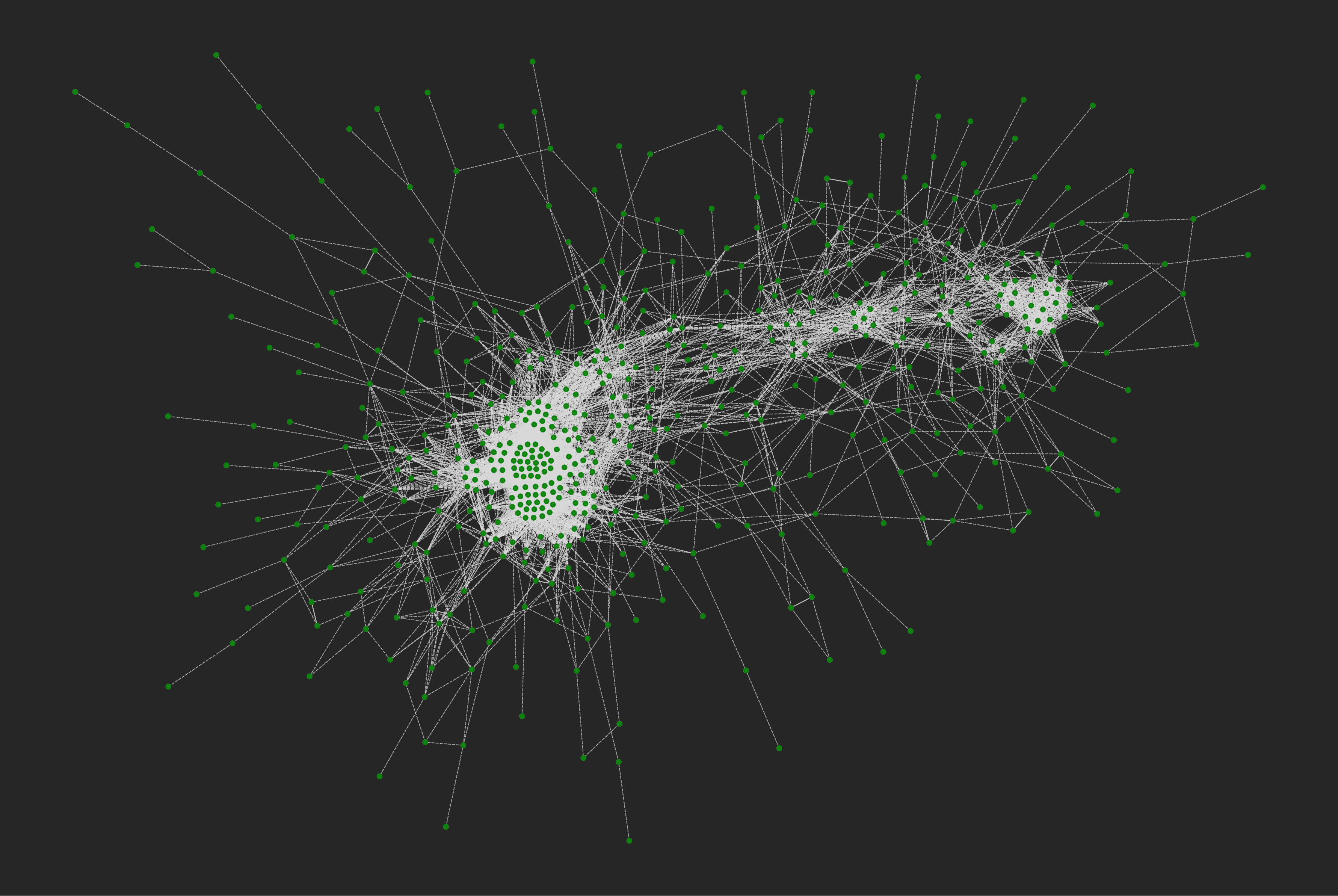} 
	\caption{The projected networks of the bipartite graph linking participants who agree on 11 of 13 responses for France (blue, top) and those who agree on 12 of 13 responses for Bangladesh (green, bottom). Each has a giant component of over 70\% of the respondants.}
	\label{fig:wellcome_participant}
\end{figure*}

This type of participant projection method is shown in fig.~\ref{fig:wellcome_participant} for France (blue nodes) and Bangladesh (green nodes). For the French participants, a giant component (capturing 70.5\% of the nodes) is formed at an agreement of 11 (of 13). In Bangladesh, a giant component forms (with 71.4\% of the nodes) at an agreement of 12. Hence, in the Bangladesh network the large clusters are groups of participants who agree on all items and each subsequent group is one edge away from that cluster. 
Without resorting to network measures, it is clear that the opinion-based group networks are quite different for the two countries. However, it should be noted that comparing networks when they have different thresholds for network formation would be not be advised, instead it would be better to compare the networks at the same threshold (11 in this case or lower if it is desired to capture more participants).

The layout algorithm used here is a force-directed layout algorithm~\cite{fruchterman1991graph} as implemented by the \emph{graph-tool} library in Python using~\cite{huefficient}. Other layout algorithms that aim to reduce the number of edges crossing will yield similar results (eg. Kamada-Kawai~\cite{kamada1989algorithm}).

\subsection{Score-based method}

A disadvantage of the previous method is, if two participants are linked it does not take into account how much they diverged. For example, in the Bangladesh projection in fig.~\ref{fig:wellcome_participant}, if, two participants differed where one has complete trust in ``Scientists'' and the other having some trust on those two items, they would not be linked in that network despite having similar opinions. Instead we introduce a \emph{score-based} linking method. This takes into account how far away two participants are on the scale provided in the survey. 

To do this, we  renormalise the scale from -1 to +1 going in equal increments corresponding to the scale. For example, if it is a 5-point scale, we take -1, -1/2, 0, 1/2, 1. On a 4-point scale this would be -1, -1/3, 1/3, 1. This value is then assigned to the edge in the bipartite graph. When we take the projection, we then compute the difference between each pair of nodes across all item responses.  To make the threshold easier to conceptualise, we subtract the number of items from the difference score 
which gives each pair of vertices a similarity score going from $-n_{\text{items}}$ to $+n_{\text{items}}$ where $n_{\text{items}}$  is the total number of items. 

\begin{figure*}
	\centering
	\includegraphics[width=0.95\textwidth]{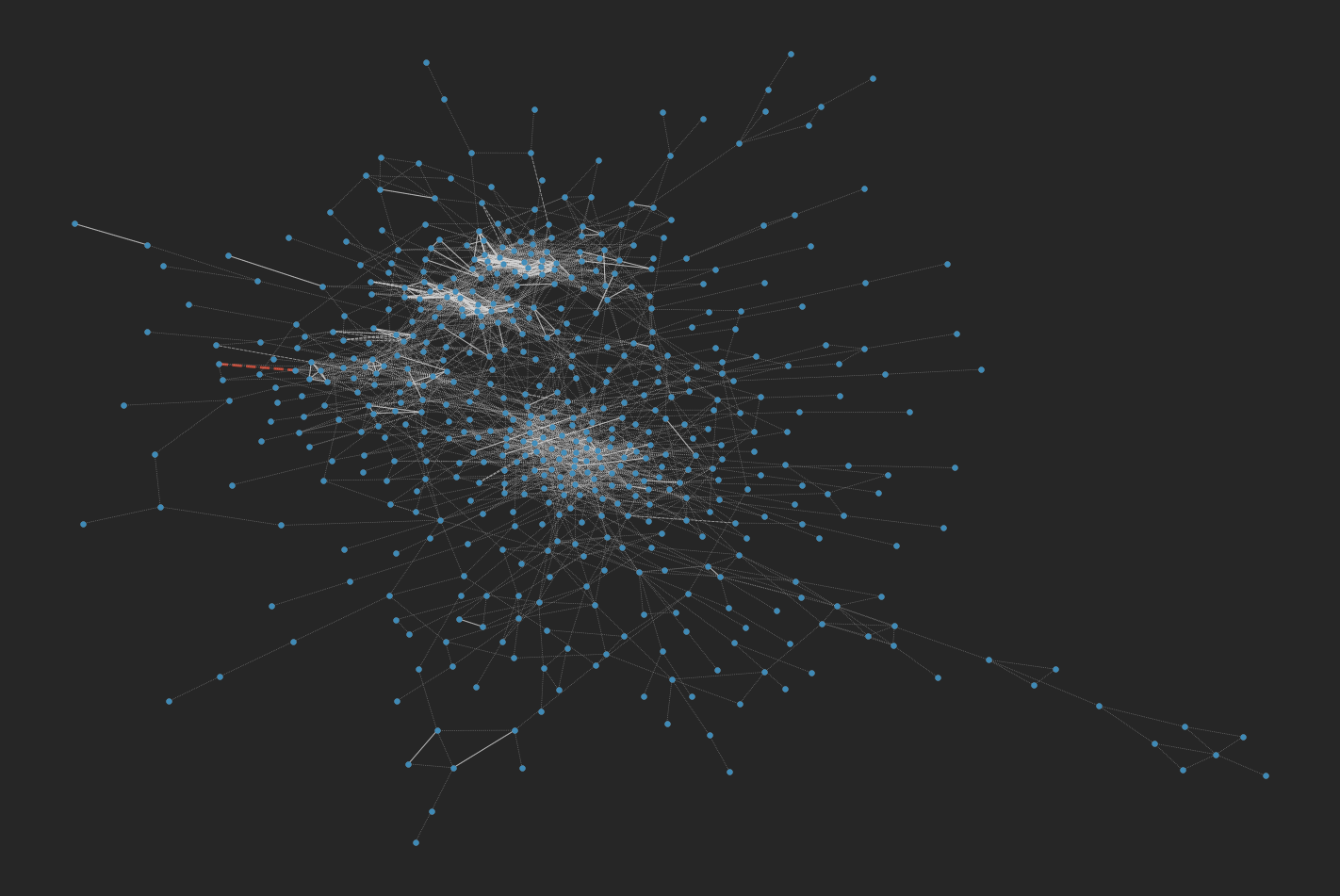} 
	\includegraphics[width=0.95\textwidth]{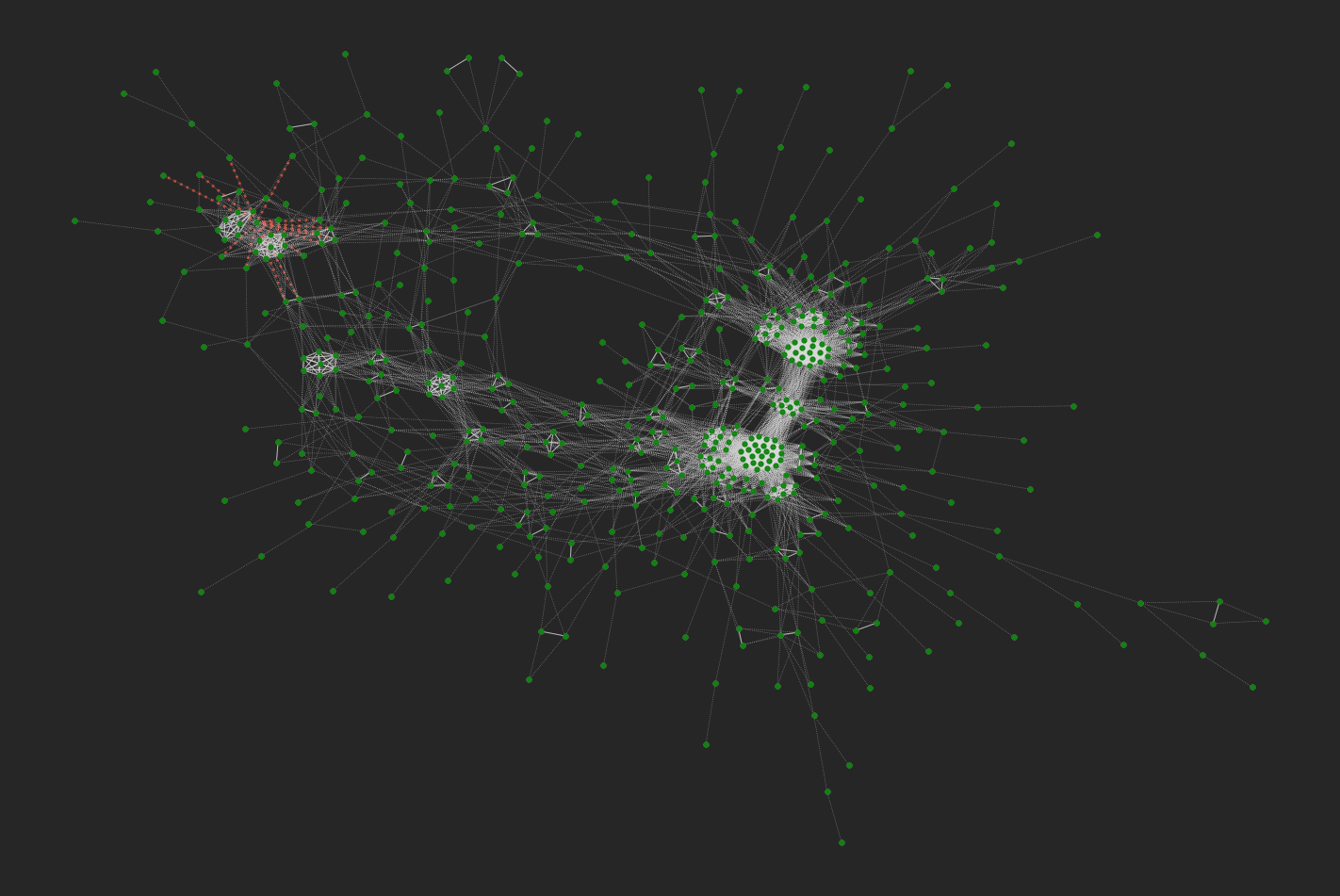} 
	\caption{The projected networks of the bipartite graph linking participants from France (blue, top) and Bangladesh (green, bottom) based on the similarity score between participants responses. Each has a giant component of over 65\% of the respondants.}
	\label{fig:wellcome_participant_score}
\end{figure*}

In the Wellcome Trust data above, with 13 items, the difference can go from 0 (i.e. they both answer the same to everything) to 26 (i.e. the are on opposite extremes for each answer). To compute the similarity between them we subtract the number of items (i.e. 13) from this difference so each pair of vertices will now have a value from -13 to +13. 

Fig.~\ref{fig:wellcome_participant_score} shows these projections for France with an agreement 
threshold of  11.5 (blue, top) and Bangladesh with an agreement threshold of 12 (green, bottom). At an agreement threshold of 12, France had a giant component of 31\%, whereas each is above 65\% in the networks displayed.
We have also included negative edges with a threshold of -3 or lower in each (on average this requires two participants being more than a value of one away for each item on the renormalised scale). In each case clusters of opinion-based groups can be identified visually. Using community detection algorithms, these could be identified and studied in more detail (see below).

Note that if a participant does not respond to an item with an extreme value, then they cannot have a difference of 2. Therefore, a score of $-n_{\text{items}}$ (e.g. -13 above) will be  rare as it will require one participant to answer fully positively to everything and another fully negatively. 

\begin{figure*}
	\centering
	\includegraphics[width=0.85\textwidth]{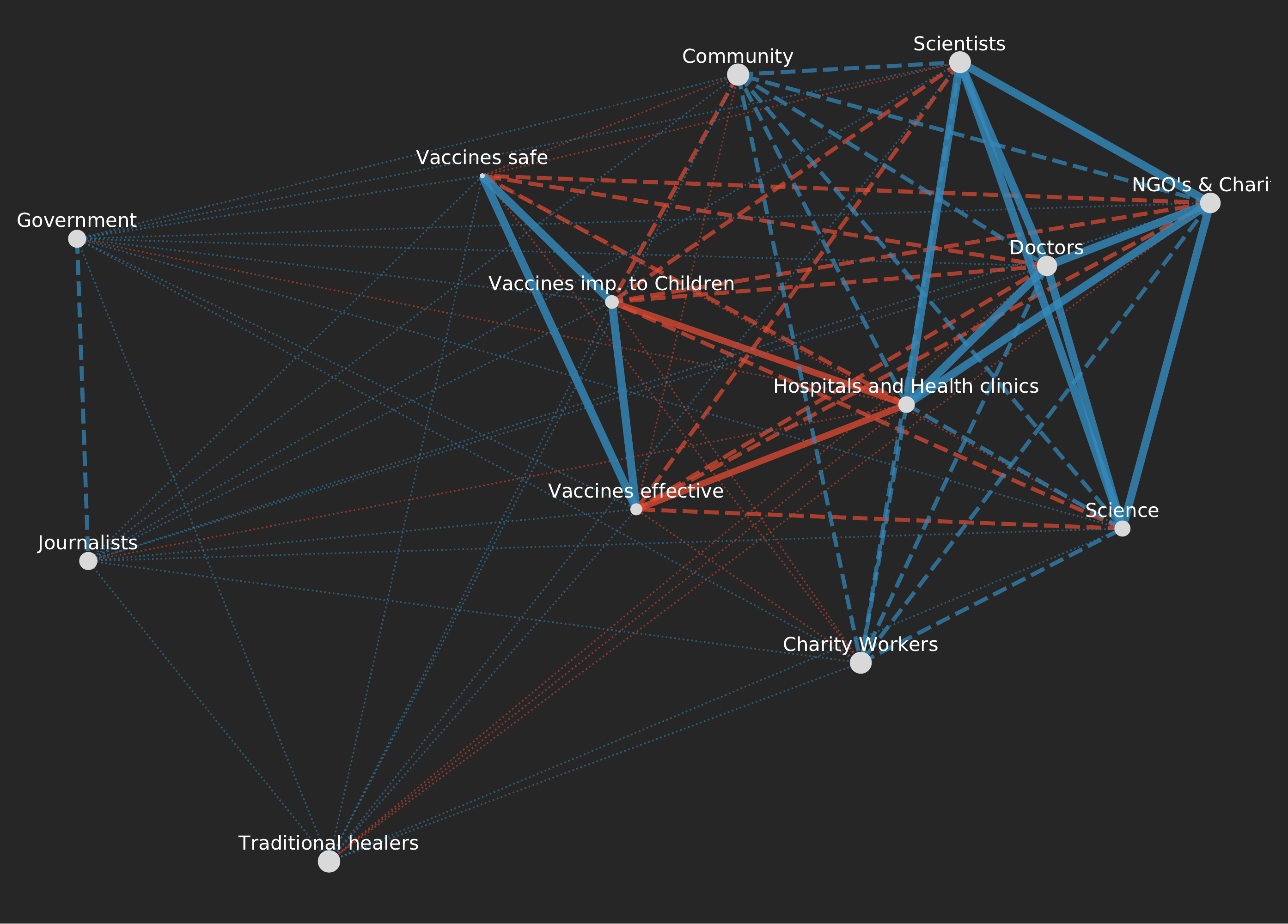} 
	\includegraphics[width=0.85\textwidth]{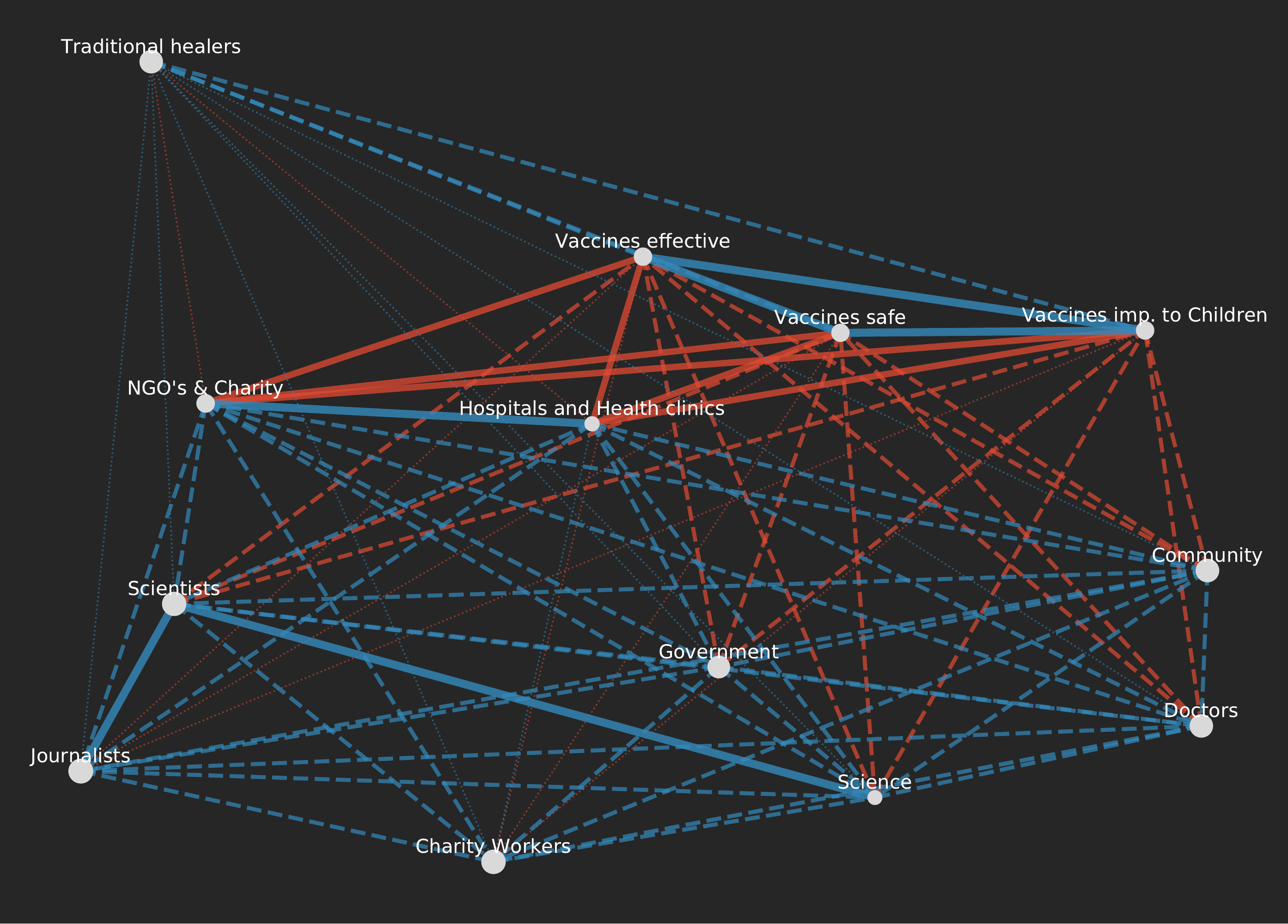} 
	\caption{The projected attitude networks of the bipartite graph linking attitudes, positively and negatively for France (top) and Bangladesh (bottom). A blue edge represents participants giving positive scores to both attitudes, and a red edge negative. The weight of each colour edge is divided into thirds.}
	\label{fig:wellcome_attitude}
\end{figure*}

A similar method can be applied to obtain a projection of links between items in what we call the attitude projection. This is shown for France and Bangladesh again in fig.~\ref{fig:wellcome_attitude}. Here, the weight between two items goes from the total number of participants $N$ to $-N$. For the purposes of visualisation, dotted blue edges are used to represent a weight of the first third, dashed represent from $N/3$ to $2N/3$ and solid for the top third. The red edges represent the same thirds but for negative values. Here we see that in France, there are a group of participants who trust the government and journalists, but are removed from the rest.


\subsection{Further examples}

\begin{figure*}
	\centering
	\includegraphics[width=0.95\textwidth]{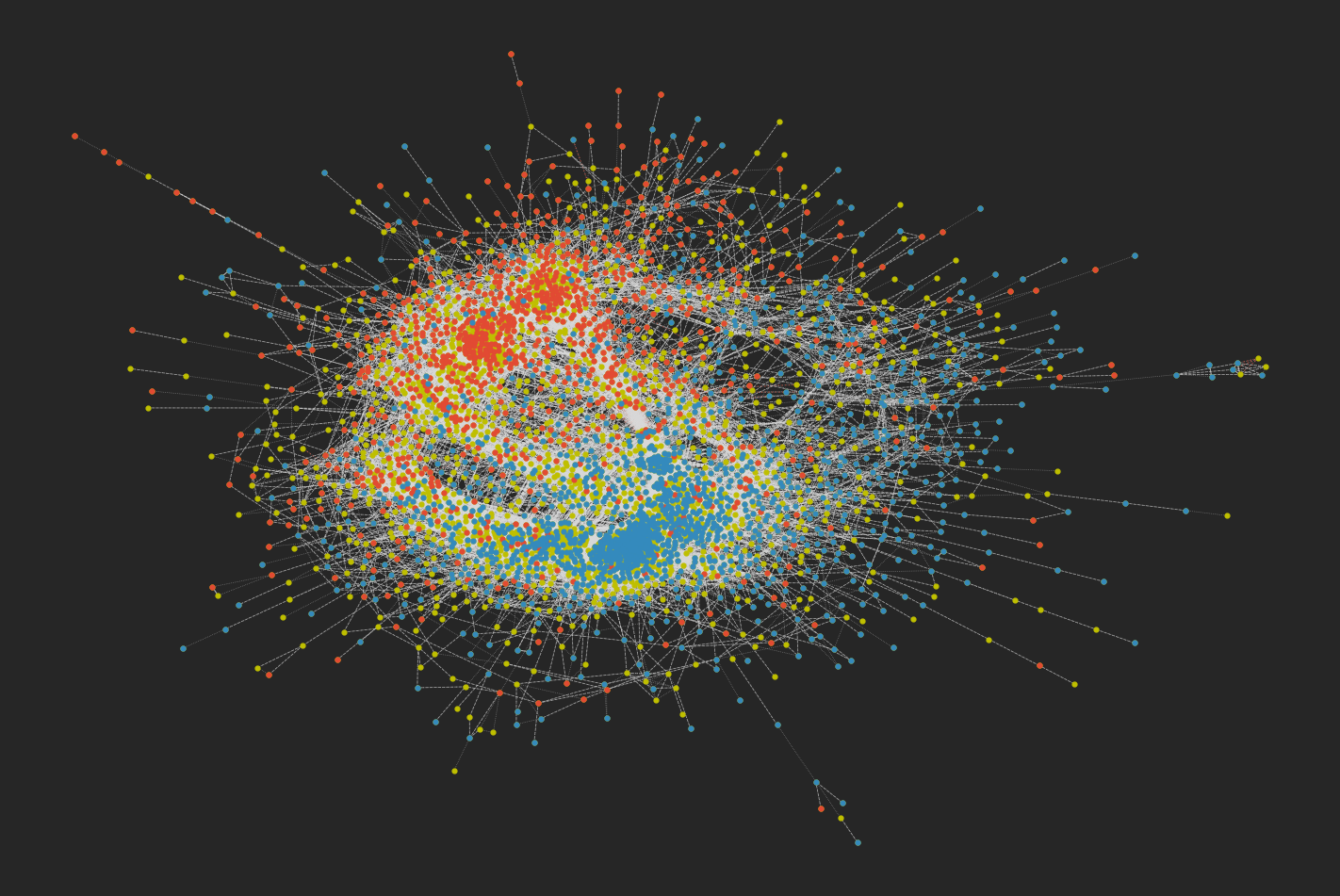} 
	\includegraphics[width=0.95\textwidth]{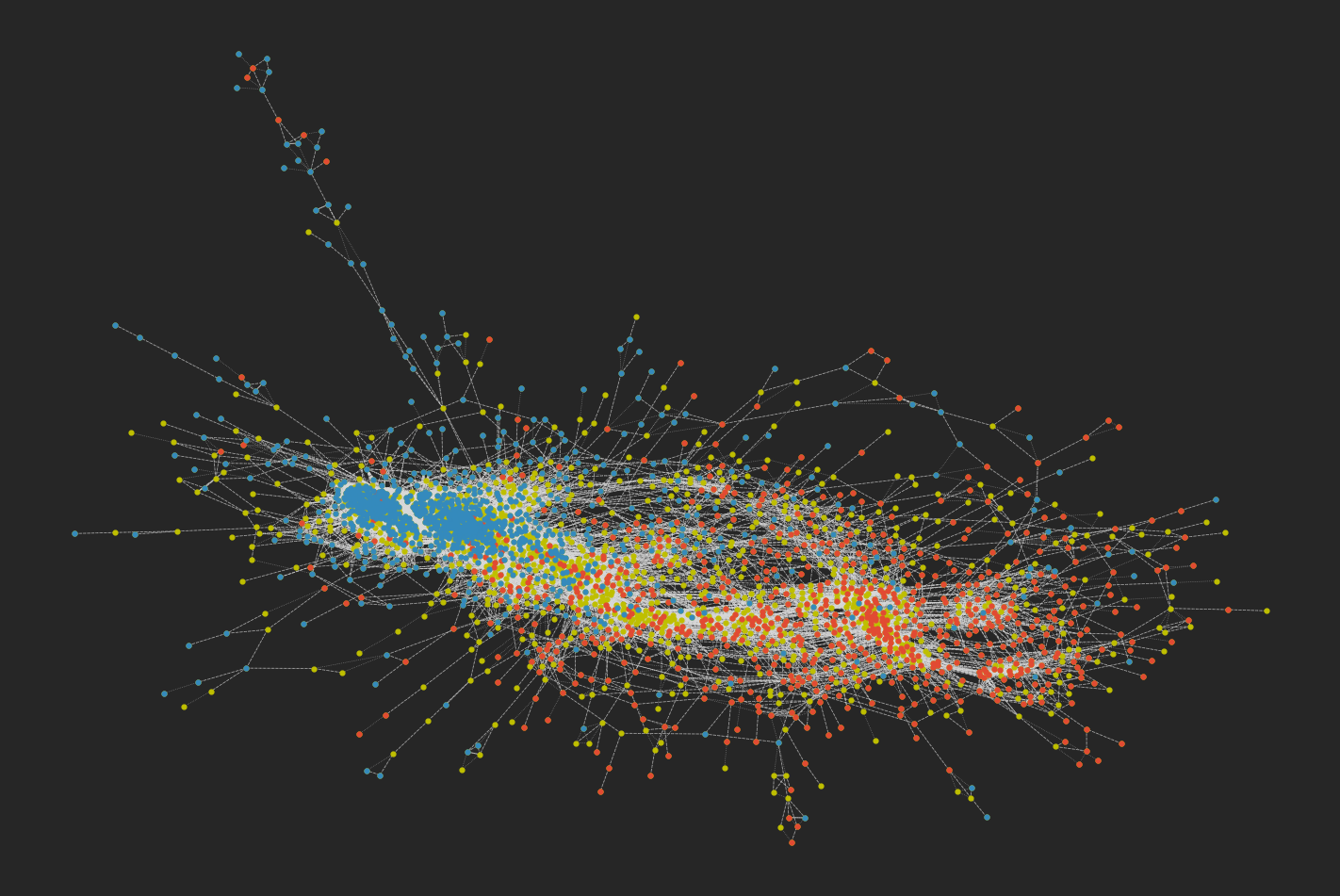} 
	\caption{The score-based participant projection for the ANES data for 2012 (top) and 2016 (bottom). Blue nodes are participants who self-identify as Democrats, red Republicans, yellow as Other/Independent. In 2016 the clusters of republicans and democrats appear further apart (i.e. there is a longer path from nodes on the extremes from one cluster to the other). }
	\label{fig:ANES_score}
\end{figure*}

In the following examples, we use our visualisation method to assess polarisation among the Unites States (US) public. To achieve this we model American National Election Studies (ANES) data. These surveys produce large representative datasets of public attitude towards politics during different US elections. For the following example, we use the 2012 and 2016 pre-election datasets and take a number of items commonly used to assess partisanship in US politics ~\cite{malka2014needs}). These are views on: abortion, income, immigration, welfare, gay marriage, business, gun control and gay marriage. In each case we apply the score based method to the participant projection and display with a threshold of 7 (maximum of 8) in fig.~\ref{fig:ANES_score}. We then coloured the nodes based on whether respondents self-identify as either democrats (blue), republicans (red) or other/independent (yellow).

The layout algorithm demonstrates a clear distinction between democrats and republicans. Using the Girvan-Newman algorithm~\cite{girvan2002community} to recursively remove nodes of high betweenness, we observe in the 2016 dataset there are 375 edges tying the two largest clusters together (this is 0.7\%) of the edges. In 2012, however, the algorithm removes smaller communities each time suggesting that either the democrats and republicans are not as easily separated, or that the clusters of democrats and republicans are not as unified so the algorithm separates sub-communities of each. Such a pattern is indicative of the growing polarisation that occurred in US society between these two time-points ~\cite{iyengar2019origins}.

\begin{figure*}
	\centering
	\includegraphics[width=0.95\textwidth]{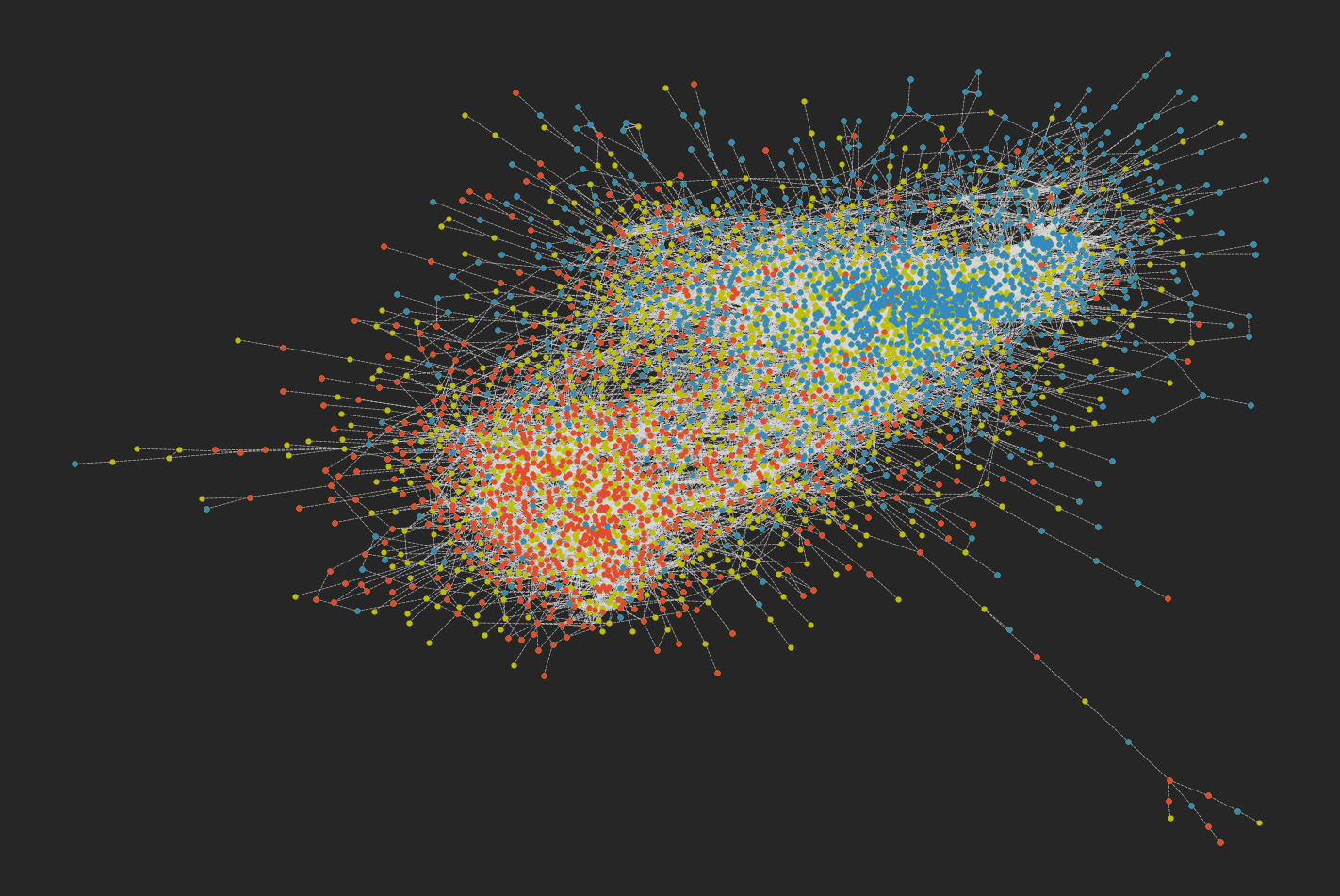} 
		\includegraphics[width=0.95\textwidth]{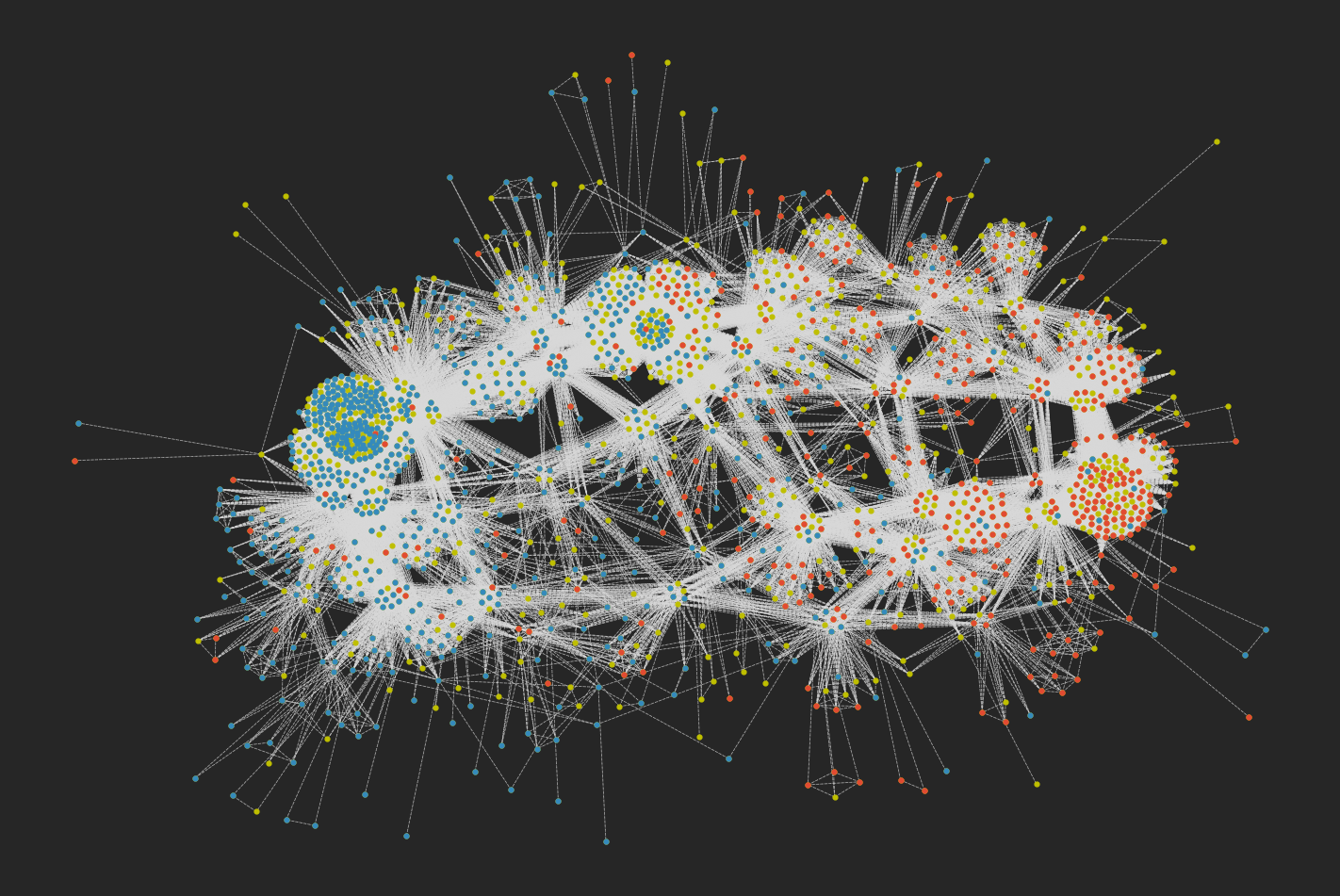}
	\caption{The participant projection for the 2012 ANES data with no scoring (top) and binarised with no score (bottom). }
	\label{fig:ANES_noscore}
\end{figure*}

In fig.~\ref{fig:ANES_noscore}, we also display the non-scoring version where we link participants based on the number of questions they answer in common for 2012. Here, we make one simplification by ``binarising'' the data. In this case, any pair that answer the same question both positively or both negatively get linked instead of having to get the same value. This is shown in the bottom of fig.~\ref{fig:ANES_noscore}. In both these figures, each cluster is separated by a distance of one attitude. Due to the simplification of the binarised version, these clusters, i.e. opinion-based groups, are much more clearly visible. Note that in both cases, the agreement score is $n_{\text{items}}-1$.

Using the Girvan-Newman algorithm, we can successfully separate two main clusters in 2012 in this case, here removing 2\% of the edges splits into two similarly sized groups, one with predominantly republicans, the other democrats.

\section{Conclusions}

In this study we have introduced a method to map opinion-based groups and attitude networks from survey data. In the case of the opinion-based groups, there is a single parameter, how much agreement one wishes the participants to have. In the case of the attitude networks, there are no real parameters, the weight of the edges tell you how often the attitudes are held together positively and negatively. 

In the participant projections, community detection methods can be used to identify polarised groups. We show that this method captures political groupings (i.e. republicans and democrats) in ANES data. Furthermore, we demonstrate that from 2012 to 2016, participants became more ideologically polarised in the same attitudes, an observation previously made in the social sciences~\cite{iyengar2019origins}. When taking other datasets, such as the Wellcome Trust Global Health Monitor data, we are similarly able to identify distinct groups of people separated on different items. 

We have previously used this agreement threshold method to map the emergence of novel opinion-based groups \cite{maher2020mapping}. In that study, survey a group of participants across three time points using the same items as the Wellcome Trust Global Health Monitor. Each survey was carried out one week apart in the UK in March 2020 during the early phase of the COVID-19 pandemic. In the final period of data collection $N=253$, we observe two attitude-based clusters separated by only 7 edges. These two groups were not observable in earlier time-points and were mostly separated by trust in science, suggesting a change in science attitudes at the onset of the COVID-19 outbreak in the UK.

With the attitude projections in fig.~\ref{fig:wellcome_attitude}, we can identify attitudes that are commonly held together. For example, in France, a large number of people hold doctors, scientists, NGOs \& charities and hospitals and health centres in a high regard, these are quite separate from people who have a lot of trust in government and journalists. In Bangladesh on the other hand, participants with a high trust in scientists also have a high trust in journalists. Similarly those who trust vaccines are more likely to trust traditional healers than in France. 

These attitude projections are similar in principle to the ones constructed in \cite{brandt2019central,dalege2016toward}. However, an advantage of our method, is the ease of computation and interpretation. Our edges represent the number of participants agreeing or disagreeing with each pair of attitudes as opposed to a partial correlation. We therefore make less theoretical and statistical assumptions about the connections between nodes in our network. 

Further network measures could be used to analyse the participant projections, however care must be taken when interpreting these, as they are not social networks. The participants in these do not know each other, they just align on certain attitudes. Therefore quantities will have to be interpreted in a different way.  
Clustering algorithms however are of particular interest to identify group-based alignment and polarisation. A next step is to use bipartite partition methods, such as ~\cite{mrvar2009partitioning} on the initial graph, and other statistical clustering methods on the raw data, to compare the opinion-based groups identified.

One issue with the stopping when the giant component is reached is the loss of some participants. If more participants  are desired the agreement threshold could be lowered and clustering algorithms relying on the weight of the edge could be used instead. Using the size of the giant component as the parameter is useful when trying to gauge if large groups are highly polarised, however, it will miss small groups of participants with shared extreme attitudes. Therefore the choice of the parameter depends on the question being asked.

Further work from a network approach is to compare the projections to random null-models, this could also be used to ascertain a significance of the edges observed which would be of particular relevance to the  clusters identified. In the case of the binarised ANES data (lower panel of fig.~\ref{fig:ANES_noscore}), there are $2^8=256$ possible clusters based on eight binarised attitudes.  However, from the clusters drawn, less than $10\%$ of these are realised in the empirical sample.

In conclusion, we present a novel and simple means of visualising factional alignment and polarisation with survey data. This is a method that allows researchers to map the emergence of opinion-based groups using standard questionnaires and panel designs. We hope this work can go some way towards facilitating research that addresses the detrimental societal consequences of polaristaion driven by opinion-based group identification.

\subsection*{List of Abbreviations}

ANES -- Americal National Election Studies

\section*{Declarations}

\noindent \textbf{Availability of data and materials}

\noindent The data used comes from secondary datasets which can be found in the relevant references.\\

\noindent \textbf{Competing interests}

\noindent The authors declare that they have no competing interests.\\

\noindent \textbf{Funding}

\noindent This project has received funding from the European Research Council (ERC) under the European Union's Horizon 2020 research and innovation programme (grant agreement No. 802421).\\

\noindent \textbf{Authors' contributions}

\noindent PMC developed the code, figures and co-wrote the manuscript, PM conceptually co-developed the model and co-wrote the manuscript and MQ conceptually co-developed the model and co-wrote the manuscript.\\

\noindent \textbf{Acknowledgements}

\noindent We would like to thank James Glesson, David O'Sullivan and Susan Fennell for discussions about the creation of the model.






\bibliography{references}

\begin{thebibliography}{33}
\expandafter\ifx\csname natexlab\endcsname\relax\def\natexlab#1{#1}\fi
\expandafter\ifx\csname bibnamefont\endcsname\relax
  \def\bibnamefont#1{#1}\fi
\expandafter\ifx\csname bibfnamefont\endcsname\relax
  \def\bibfnamefont#1{#1}\fi
\expandafter\ifx\csname citenamefont\endcsname\relax
  \def\citenamefont#1{#1}\fi
\expandafter\ifx\csname url\endcsname\relax
  \def\url#1{\texttt{#1}}\fi
\expandafter\ifx\csname urlprefix\endcsname\relax\def\urlprefix{URL }\fi
\providecommand{\bibinfo}[2]{#2}
\providecommand{\eprint}[2][]{\url{#2}}

\bibitem[{\citenamefont{Bliuc et~al.}(2007)\citenamefont{Bliuc, McGarty,
  Reynolds, and Muntele}}]{bliuc2007opinion}
\bibinfo{author}{\bibfnamefont{A.-M.} \bibnamefont{Bliuc}},
  \bibinfo{author}{\bibfnamefont{C.}~\bibnamefont{McGarty}},
  \bibinfo{author}{\bibfnamefont{K.}~\bibnamefont{Reynolds}}, \bibnamefont{and}
  \bibinfo{author}{\bibfnamefont{D.}~\bibnamefont{Muntele}},
  \bibinfo{journal}{European Journal of Social Psychology}
  \textbf{\bibinfo{volume}{37}}, \bibinfo{pages}{19} (\bibinfo{year}{2007}).

\bibitem[{\citenamefont{McGarty et~al.}(2009)\citenamefont{McGarty, Bliuc,
  Thomas, and Bongiorno}}]{mcgarty2009collective}
\bibinfo{author}{\bibfnamefont{C.}~\bibnamefont{McGarty}},
  \bibinfo{author}{\bibfnamefont{A.-M.} \bibnamefont{Bliuc}},
  \bibinfo{author}{\bibfnamefont{E.~F.} \bibnamefont{Thomas}},
  \bibnamefont{and}
  \bibinfo{author}{\bibfnamefont{R.}~\bibnamefont{Bongiorno}},
  \bibinfo{journal}{Journal of Social Issues} \textbf{\bibinfo{volume}{65}},
  \bibinfo{pages}{839} (\bibinfo{year}{2009}).

\bibitem[{\citenamefont{Hogg and Smith}(2007)}]{Hogg2007a}
\bibinfo{author}{\bibfnamefont{M.~A.} \bibnamefont{Hogg}} \bibnamefont{and}
  \bibinfo{author}{\bibfnamefont{J.~R.} \bibnamefont{Smith}},
  \bibinfo{journal}{European Review of Social Psychology}
  \textbf{\bibinfo{volume}{18}}, \bibinfo{pages}{89} (\bibinfo{year}{2007}).

\bibitem[{\citenamefont{Bliuc et~al.}(2015)\citenamefont{Bliuc, McGarty,
  Thomas, Lala, Berndsen, and Misajon}}]{Bliuc2015}
\bibinfo{author}{\bibfnamefont{A.-M.} \bibnamefont{Bliuc}},
  \bibinfo{author}{\bibfnamefont{C.}~\bibnamefont{McGarty}},
  \bibinfo{author}{\bibfnamefont{E.~F.} \bibnamefont{Thomas}},
  \bibinfo{author}{\bibfnamefont{G.}~\bibnamefont{Lala}},
  \bibinfo{author}{\bibfnamefont{M.}~\bibnamefont{Berndsen}}, \bibnamefont{and}
  \bibinfo{author}{\bibfnamefont{R.}~\bibnamefont{Misajon}},
  \bibinfo{journal}{Nature Climate Change} \textbf{\bibinfo{volume}{5}},
  \bibinfo{pages}{226} (\bibinfo{year}{2015}),
  \urlprefix\url{https://doi.org/10.1038/nclimate2507}.

\bibitem[{\citenamefont{Maher et~al.}(2020)\citenamefont{Maher, MacCarron, and
  Quayle}}]{maher2020mapping}
\bibinfo{author}{\bibfnamefont{P.~J.} \bibnamefont{Maher}},
  \bibinfo{author}{\bibfnamefont{P.}~\bibnamefont{MacCarron}},
  \bibnamefont{and} \bibinfo{author}{\bibfnamefont{M.}~\bibnamefont{Quayle}},
  \bibinfo{journal}{British Journal of Social Psychology}
  \textbf{\bibinfo{volume}{59}}, \bibinfo{pages}{641} (\bibinfo{year}{2020}).

\bibitem[{\citenamefont{Van~Bavel et~al.}(2020)\citenamefont{Van~Bavel,
  Baicker, Boggio, Capraro, Cichocka, Cikara, Crockett, Crum, Douglas, Druckman
  et~al.}}]{van2020using}
\bibinfo{author}{\bibfnamefont{J.~J.} \bibnamefont{Van~Bavel}},
  \bibinfo{author}{\bibfnamefont{K.}~\bibnamefont{Baicker}},
  \bibinfo{author}{\bibfnamefont{P.~S.} \bibnamefont{Boggio}},
  \bibinfo{author}{\bibfnamefont{V.}~\bibnamefont{Capraro}},
  \bibinfo{author}{\bibfnamefont{A.}~\bibnamefont{Cichocka}},
  \bibinfo{author}{\bibfnamefont{M.}~\bibnamefont{Cikara}},
  \bibinfo{author}{\bibfnamefont{M.~J.} \bibnamefont{Crockett}},
  \bibinfo{author}{\bibfnamefont{A.~J.} \bibnamefont{Crum}},
  \bibinfo{author}{\bibfnamefont{K.~M.} \bibnamefont{Douglas}},
  \bibinfo{author}{\bibfnamefont{J.~N.} \bibnamefont{Druckman}},
  \bibnamefont{et~al.}, \bibinfo{journal}{Nature Human Behaviour} pp.
  \bibinfo{pages}{1--12} (\bibinfo{year}{2020}).

\bibitem[{\citenamefont{Garcia et~al.}(2019)\citenamefont{Garcia, Galaz, and
  Daume}}]{garcia2019eatlancet}
\bibinfo{author}{\bibfnamefont{D.}~\bibnamefont{Garcia}},
  \bibinfo{author}{\bibfnamefont{V.}~\bibnamefont{Galaz}}, \bibnamefont{and}
  \bibinfo{author}{\bibfnamefont{S.}~\bibnamefont{Daume}},
  \bibinfo{journal}{The Lancet} \textbf{\bibinfo{volume}{394}},
  \bibinfo{pages}{2153} (\bibinfo{year}{2019}).

\bibitem[{\citenamefont{Macy et~al.}(2019)\citenamefont{Macy, Deri, Ruch, and
  Tong}}]{macy2019opinion}
\bibinfo{author}{\bibfnamefont{M.}~\bibnamefont{Macy}},
  \bibinfo{author}{\bibfnamefont{S.}~\bibnamefont{Deri}},
  \bibinfo{author}{\bibfnamefont{A.}~\bibnamefont{Ruch}}, \bibnamefont{and}
  \bibinfo{author}{\bibfnamefont{N.}~\bibnamefont{Tong}},
  \bibinfo{journal}{Science advances} \textbf{\bibinfo{volume}{5}},
  \bibinfo{pages}{eaax0754} (\bibinfo{year}{2019}).

\bibitem[{\citenamefont{McPherson et~al.}(2001)\citenamefont{McPherson,
  Smith-Lovin, and Cook}}]{mcpherson2001birds}
\bibinfo{author}{\bibfnamefont{M.}~\bibnamefont{McPherson}},
  \bibinfo{author}{\bibfnamefont{L.}~\bibnamefont{Smith-Lovin}},
  \bibnamefont{and} \bibinfo{author}{\bibfnamefont{J.~M.} \bibnamefont{Cook}},
  \bibinfo{journal}{Annual review of sociology} \textbf{\bibinfo{volume}{27}},
  \bibinfo{pages}{415} (\bibinfo{year}{2001}).

\bibitem[{\citenamefont{Jetten et~al.}(1997)\citenamefont{Jetten, Spears, and
  Manstead}}]{jetten1997strength}
\bibinfo{author}{\bibfnamefont{J.}~\bibnamefont{Jetten}},
  \bibinfo{author}{\bibfnamefont{R.}~\bibnamefont{Spears}}, \bibnamefont{and}
  \bibinfo{author}{\bibfnamefont{A.~S.} \bibnamefont{Manstead}},
  \bibinfo{journal}{European journal of social psychology}
  \textbf{\bibinfo{volume}{27}}, \bibinfo{pages}{603} (\bibinfo{year}{1997}).

\bibitem[{\citenamefont{Bakshy et~al.}(2015)\citenamefont{Bakshy, Messing, and
  Adamic}}]{bakshy2015exposure}
\bibinfo{author}{\bibfnamefont{E.}~\bibnamefont{Bakshy}},
  \bibinfo{author}{\bibfnamefont{S.}~\bibnamefont{Messing}}, \bibnamefont{and}
  \bibinfo{author}{\bibfnamefont{L.~A.} \bibnamefont{Adamic}},
  \bibinfo{journal}{Science} \textbf{\bibinfo{volume}{348}},
  \bibinfo{pages}{1130} (\bibinfo{year}{2015}).

\bibitem[{\citenamefont{B{\'e}langer et~al.}(2020)\citenamefont{B{\'e}langer,
  Schumpe, Nisa, and Moyano}}]{belanger2020countermessaging}
\bibinfo{author}{\bibfnamefont{J.~J.} \bibnamefont{B{\'e}langer}},
  \bibinfo{author}{\bibfnamefont{B.~M.} \bibnamefont{Schumpe}},
  \bibinfo{author}{\bibfnamefont{C.~F.} \bibnamefont{Nisa}}, \bibnamefont{and}
  \bibinfo{author}{\bibfnamefont{M.}~\bibnamefont{Moyano}},
  \bibinfo{journal}{Motivation Science}  (\bibinfo{year}{2020}).

\bibitem[{\citenamefont{Axelrod}(1997)}]{axelrod1997dissemination}
\bibinfo{author}{\bibfnamefont{R.}~\bibnamefont{Axelrod}},
  \bibinfo{journal}{Journal of conflict resolution}
  \textbf{\bibinfo{volume}{41}}, \bibinfo{pages}{203} (\bibinfo{year}{1997}).

\bibitem[{\citenamefont{Valori et~al.}(2012)\citenamefont{Valori, Picciolo,
  Allansdottir, and Garlaschelli}}]{valori2012reconciling}
\bibinfo{author}{\bibfnamefont{L.}~\bibnamefont{Valori}},
  \bibinfo{author}{\bibfnamefont{F.}~\bibnamefont{Picciolo}},
  \bibinfo{author}{\bibfnamefont{A.}~\bibnamefont{Allansdottir}},
  \bibnamefont{and}
  \bibinfo{author}{\bibfnamefont{D.}~\bibnamefont{Garlaschelli}},
  \bibinfo{journal}{Proceedings of the National Academy of Sciences}
  \textbf{\bibinfo{volume}{109}}, \bibinfo{pages}{1068} (\bibinfo{year}{2012}).

\bibitem[{\citenamefont{Breiger et~al.}(2014)\citenamefont{Breiger, Schoon,
  Melamed, Asal, and Rethemeyer}}]{breiger2014comparative}
\bibinfo{author}{\bibfnamefont{R.~L.} \bibnamefont{Breiger}},
  \bibinfo{author}{\bibfnamefont{E.}~\bibnamefont{Schoon}},
  \bibinfo{author}{\bibfnamefont{D.}~\bibnamefont{Melamed}},
  \bibinfo{author}{\bibfnamefont{V.}~\bibnamefont{Asal}}, \bibnamefont{and}
  \bibinfo{author}{\bibfnamefont{R.~K.} \bibnamefont{Rethemeyer}},
  \bibinfo{journal}{Social Networks} \textbf{\bibinfo{volume}{36}},
  \bibinfo{pages}{23} (\bibinfo{year}{2014}).

\bibitem[{\citenamefont{B{\u{a}}beanu et~al.}(2018)\citenamefont{B{\u{a}}beanu,
  van~de Vis, and Garlaschelli}}]{buabeanu2018ultrametricity}
\bibinfo{author}{\bibfnamefont{A.-I.} \bibnamefont{B{\u{a}}beanu}},
  \bibinfo{author}{\bibfnamefont{J.}~\bibnamefont{van~de Vis}},
  \bibnamefont{and}
  \bibinfo{author}{\bibfnamefont{D.}~\bibnamefont{Garlaschelli}},
  \bibinfo{journal}{New Journal of Physics} \textbf{\bibinfo{volume}{20}},
  \bibinfo{pages}{103026} (\bibinfo{year}{2018}).

\bibitem[{\citenamefont{Erickson}(1988)}]{erickson1988relational}
\bibinfo{author}{\bibfnamefont{B.~H.} \bibnamefont{Erickson}},
  \bibinfo{journal}{Social structures: A network approach}
  \textbf{\bibinfo{volume}{99}}, \bibinfo{pages}{443} (\bibinfo{year}{1988}).

\bibitem[{\citenamefont{Boutyline and Vaisey}(2017)}]{boutyline2017belief}
\bibinfo{author}{\bibfnamefont{A.}~\bibnamefont{Boutyline}} \bibnamefont{and}
  \bibinfo{author}{\bibfnamefont{S.}~\bibnamefont{Vaisey}},
  \bibinfo{journal}{American journal of sociology}
  \textbf{\bibinfo{volume}{122}}, \bibinfo{pages}{1371} (\bibinfo{year}{2017}).

\bibitem[{\citenamefont{Brandt et~al.}(2019)\citenamefont{Brandt, Sibley, and
  Osborne}}]{brandt2019central}
\bibinfo{author}{\bibfnamefont{M.~J.} \bibnamefont{Brandt}},
  \bibinfo{author}{\bibfnamefont{C.~G.} \bibnamefont{Sibley}},
  \bibnamefont{and} \bibinfo{author}{\bibfnamefont{D.}~\bibnamefont{Osborne}},
  \bibinfo{journal}{Personality and Social Psychology Bulletin}
  \textbf{\bibinfo{volume}{45}}, \bibinfo{pages}{1352} (\bibinfo{year}{2019}).

\bibitem[{\citenamefont{DeVellis}(2003)}]{DeVellis2003}
\bibinfo{author}{\bibfnamefont{R.}~\bibnamefont{DeVellis}},
  \emph{\bibinfo{title}{{Scale development: theory and applications}}}, Applied
  social research methods series (\bibinfo{publisher}{Thousand Oaks: Sage
  Publications, Inc.}, \bibinfo{year}{2003}), ISBN
  \bibinfo{isbn}{9780761926054}.

\bibitem[{\citenamefont{Breiger}(1974)}]{breiger1974duality}
\bibinfo{author}{\bibfnamefont{R.~L.} \bibnamefont{Breiger}},
  \bibinfo{journal}{Social forces} \textbf{\bibinfo{volume}{53}},
  \bibinfo{pages}{181} (\bibinfo{year}{1974}).

\bibitem[{\citenamefont{Horv\'at and Zweig}(2013)}]{Horvat2013}
\bibinfo{author}{\bibfnamefont{E.-A.} \bibnamefont{Horv\'at}} \bibnamefont{and}
  \bibinfo{author}{\bibfnamefont{K.~A.} \bibnamefont{Zweig}},
  \bibinfo{journal}{Social Network Analysis and Mining}
  \textbf{\bibinfo{volume}{3}}, \bibinfo{pages}{1209} (\bibinfo{year}{2013}),
  ISSN \bibinfo{issn}{1869-5469},
  \urlprefix\url{https://doi.org/10.1007/s13278-013-0133-9}.

\bibitem[{\citenamefont{Dalege et~al.}(2016)\citenamefont{Dalege, Borsboom, van
  Harreveld, van~den Berg, Conner, and van~der Maas}}]{dalege2016toward}
\bibinfo{author}{\bibfnamefont{J.}~\bibnamefont{Dalege}},
  \bibinfo{author}{\bibfnamefont{D.}~\bibnamefont{Borsboom}},
  \bibinfo{author}{\bibfnamefont{F.}~\bibnamefont{van Harreveld}},
  \bibinfo{author}{\bibfnamefont{H.}~\bibnamefont{van~den Berg}},
  \bibinfo{author}{\bibfnamefont{M.}~\bibnamefont{Conner}}, \bibnamefont{and}
  \bibinfo{author}{\bibfnamefont{H.~L.} \bibnamefont{van~der Maas}},
  \bibinfo{journal}{Psychological review} \textbf{\bibinfo{volume}{123}},
  \bibinfo{pages}{2} (\bibinfo{year}{2016}).

\bibitem[{\citenamefont{Mrvar and Doreian}(2009)}]{mrvar2009partitioning}
\bibinfo{author}{\bibfnamefont{A.}~\bibnamefont{Mrvar}} \bibnamefont{and}
  \bibinfo{author}{\bibfnamefont{P.}~\bibnamefont{Doreian}},
  \bibinfo{journal}{Journal of Mathematical Sociology}
  \textbf{\bibinfo{volume}{33}}, \bibinfo{pages}{196} (\bibinfo{year}{2009}).

\bibitem[{\citenamefont{Wyse et~al.}(2014)\citenamefont{Wyse, Friel, and
  Latouche}}]{wyse2014inferring}
\bibinfo{author}{\bibfnamefont{J.}~\bibnamefont{Wyse}},
  \bibinfo{author}{\bibfnamefont{N.}~\bibnamefont{Friel}}, \bibnamefont{and}
  \bibinfo{author}{\bibfnamefont{P.}~\bibnamefont{Latouche}},
  \bibinfo{journal}{arXiv preprint arXiv:1404.2911}  (\bibinfo{year}{2014}).

\bibitem[{\citenamefont{Monitor}(2018)}]{monitor2018does}
\bibinfo{author}{\bibfnamefont{W.~G.} \bibnamefont{Monitor}},
  \emph{\bibinfo{title}{How does the world feel about science and health}}
  (\bibinfo{year}{2018}).

\bibitem[{\citenamefont{Gleiser}(2007)}]{gleiser2007become}
\bibinfo{author}{\bibfnamefont{P.~M.} \bibnamefont{Gleiser}},
  \bibinfo{journal}{Journal of Statistical Mechanics: Theory and Experiment}
  \textbf{\bibinfo{volume}{2007}}, \bibinfo{pages}{P09020}
  (\bibinfo{year}{2007}).

\bibitem[{\citenamefont{Fruchterman and Reingold}(1991)}]{fruchterman1991graph}
\bibinfo{author}{\bibfnamefont{T.~M.} \bibnamefont{Fruchterman}}
  \bibnamefont{and} \bibinfo{author}{\bibfnamefont{E.~M.}
  \bibnamefont{Reingold}}, \bibinfo{journal}{Software: Practice and experience}
  \textbf{\bibinfo{volume}{21}}, \bibinfo{pages}{1129} (\bibinfo{year}{1991}).

\bibitem[{\citenamefont{Hu}(2005)}]{huefficient}
\bibinfo{author}{\bibfnamefont{Y.}~\bibnamefont{Hu}},
  \bibinfo{journal}{Mathematica Journal} \textbf{\bibinfo{volume}{10}},
  \bibinfo{pages}{37} (\bibinfo{year}{2005}).

\bibitem[{\citenamefont{Kamada et~al.}(1989)\citenamefont{Kamada, Kawai
  et~al.}}]{kamada1989algorithm}
\bibinfo{author}{\bibfnamefont{T.}~\bibnamefont{Kamada}},
  \bibinfo{author}{\bibfnamefont{S.}~\bibnamefont{Kawai}},
  \bibnamefont{et~al.}, \bibinfo{journal}{Information processing letters}
  \textbf{\bibinfo{volume}{31}}, \bibinfo{pages}{7} (\bibinfo{year}{1989}).

\bibitem[{\citenamefont{Malka et~al.}(2014)\citenamefont{Malka, Soto, Inzlicht,
  and Lelkes}}]{malka2014needs}
\bibinfo{author}{\bibfnamefont{A.}~\bibnamefont{Malka}},
  \bibinfo{author}{\bibfnamefont{C.~J.} \bibnamefont{Soto}},
  \bibinfo{author}{\bibfnamefont{M.}~\bibnamefont{Inzlicht}}, \bibnamefont{and}
  \bibinfo{author}{\bibfnamefont{Y.}~\bibnamefont{Lelkes}},
  \bibinfo{journal}{Journal of Personality and Social Psychology}
  \textbf{\bibinfo{volume}{106}}, \bibinfo{pages}{1031} (\bibinfo{year}{2014}).

\bibitem[{\citenamefont{Girvan and Newman}(2002)}]{girvan2002community}
\bibinfo{author}{\bibfnamefont{M.}~\bibnamefont{Girvan}} \bibnamefont{and}
  \bibinfo{author}{\bibfnamefont{M.~E.} \bibnamefont{Newman}},
  \bibinfo{journal}{Proceedings of the national academy of sciences}
  \textbf{\bibinfo{volume}{99}}, \bibinfo{pages}{7821} (\bibinfo{year}{2002}).

\bibitem[{\citenamefont{Iyengar et~al.}(2019)\citenamefont{Iyengar, Lelkes,
  Levendusky, Malhotra, and Westwood}}]{iyengar2019origins}
\bibinfo{author}{\bibfnamefont{S.}~\bibnamefont{Iyengar}},
  \bibinfo{author}{\bibfnamefont{Y.}~\bibnamefont{Lelkes}},
  \bibinfo{author}{\bibfnamefont{M.}~\bibnamefont{Levendusky}},
  \bibinfo{author}{\bibfnamefont{N.}~\bibnamefont{Malhotra}}, \bibnamefont{and}
  \bibinfo{author}{\bibfnamefont{S.~J.} \bibnamefont{Westwood}},
  \bibinfo{journal}{Annual Review of Political Science}
  \textbf{\bibinfo{volume}{22}}, \bibinfo{pages}{129} (\bibinfo{year}{2019}).

\end{thebibliography}
\end{document}